\documentclass[aps,pra,showpacs,twocolumn]{revtex4}
\usepackage{amssymb}
\usepackage{graphicx}
\usepackage{colordvi}
\usepackage{color}
\usepackage{dcolumn,amsmath,amsthm,amscd,amsfonts,amssymb,epsfig,graphics,graphicx,eucal}

\begin{document}

\title{Asymmetric transmission of surface plasmon polaritons on planar gratings}
\author{{V. Kuzmiak$^{1}$} and A. A. Maradudin$^2$}
\affiliation{ $^1$Institute of Photonics and Electronics, Czech Academy of Sciences, v.v.i,
Chaberska 57, 182 51 Praha 8, Czech Republic
\\
$^2$ Department of Physics and Astronomy, University of California,
 Irvine, California 92697, U.S.A.}

\date{\today}

\begin{abstract}

 We describe a surface structure consisting of a metal-air interface where the metallic part consists of two metallic segments with a periodic modulation of the interface between them. Such a structure possesses a different transmissivity for a surface plasmon polariton incident on it from one side of it than it has for a surface plasmon polariton incident on it from the opposite side. This asymmetric transmission of a surface plasmon polariton is based on the suppression of the zero-order Bragg beam which, for a certain value of the modulation depth, is not transmitted through the structure, while the diffraction efficiencies of the +1 and -1 Bragg beams can be modified by varying the period of grating and/or the angle of incidence. For a certain range of the incidence angle one can observe asymmetry in transmittance for the -1 mode while the +1 mode is completely suppressed. By varying the material and geometrical parameters of the diffractive structure one can control the contrast transmission that characterizes the degree of the asymmetry. This property of the structure is demonstrated by the results of computer simulation calculations.

\end{abstract}
\maketitle
\input{epsf.tex}
\epsfverbosetrue

\section{Introduction}
\label{intro}

In recent years there has been growing interest in investigations of two-dimensional nanostructured metallic structures. The metallic nanostrutures derive their unique optical properties from their ability to support collective electron excitations, known as surface plasmon-polaritons(SPPs). Surface plasmon polaritons are quasi-two-dimensional electromagnetic waves that propagate along a dielectric-metal interface with amplitudes that decay exponentially with increasing distance into both of the neighboring media~\cite{Agranovich}. The possibility to confine the light into sub-wavelength volumes, which stems from the latter feature, has a profound effect on the efficiency of many optical processes and makes surface plasmon-polaritons very sensitive to surface properties.
In this paper we are interested in exploring nanostructured two-dimensional metallic surfaces which may provide a platform for a realistic optical analog of one-way electronic devices such as diodes and transistors. The majority of the devices supporting unidirectional propagation of surface plasmon polaritons are based  on nonlinear optics and magneto-optical(MO) effects~\cite{Haldane}-~\cite{KEV}.
 For example, a waveguide has been designed in the form of a gap between a semi-infinite dielectric photonic crystal and a semi-infinite metal to which a static magnetic field is applied, in which electromagnetic waves can propagate in only one direction~\cite{Yu}. It was subsequently shown~\cite{KEV} that if the photonic crystal in this waveguide structure is fabricated from a transparent dielectric magneto-optic material, to which the magnetic field is applied, the window of the frequencies within which the waveguide displays one-way propagation can be achieved at much lower magnetic field strengths than are required for this purpose in the structure proposed in Ref.~\onlinecite{Yu}.

The application of a magnetic field to a structure to produce one-way propagation of the surface or guided waves it supports may not always be an option for some applications of those waves. This consideration stimulates searches for surface structures that produce one-way propagation of a surface or guided wave without the need of a magnetic field.
For example, we have shown recently~\cite{KM} that a 2D system consisting of a square array of scatterers deposited on a metal surface in a triangular mesh exhibits asymmetric transmission of a SPP when a diffractive structure is added to one side of the structure. This structure does not require either electrical nonlinearity or the presence of the magnetic field to accomplish this. The asymmetric transmission is a consequence solely of the geometry of the structure.



In this paper we describe yet another surface structure that has different transmissivities for surface plasmon polaritons incident on it from opposite directions. The surface structure consists of a metal-air interface where the metallic region is formed by two metallic segments whose interface between them is periodically modulated. We first employ a theoretical approach based on the thin phase screen model~\cite{thin_phase} which allows determining the transmitted electric field in the form of a Fourier expansion. We show that when the interface has the form of a rectangular grating with a critical value of the modulation depth, the zero-order term in the Fourier expansion of the transmitted electric field can be suppressed in a certain frequency range. The  +1 and -1 modes which unlike the zero-order beam do not satisfy reciprocity, remain propagating in this frequency range, and their diffraction efficiencies can be independently modified by varying the period of the grating and/or the angle of incidence and, as a result, the transmittance of SPP propagating through this structure may become asymmetric.

The suppression of the zero-order transmitted beam has been used effectively by Serebryannikov and his colleauges in designing structures that produce asymmetric transmission of volume electromagnetic waves through them ~\cite{Se1}-~\cite{Se3}. These are all slabs whose two surfaces are both periodically corrugated, but with different periods, and in some cases also pierced by a slit of subwavelength width. Although the structures studied in these papers are volume structures, not surface structures; the electromagnetic waves illuminating them are volume waves, not surface waves; and the means for suppressing the zero-order transmitted beam are different from ours, the mechanisms by which they produce asymmetric transmission are basically the same as those producing this effect in our surface plasmon polaritonic structure.

The predictions obtained by our use of the thin phase screen model have been verified by using numerical simulations based on the finite-element frequency-domain method, which confirm that suppression of the zero-order mode occurs when there exists a sufficiently large refractive index contrast between the two metals at a certain frequency. Consequently, the transmittance of the structure relies on the diffraction efficiencies of the +1 and -1 Bragg beams that can be modified by varying the angle of incidence. Namely, we find that within a certain range of the incidence angle an asymmetry in transmittance for the -1 mode exists, while the +1 mode is completely suppressed. By varying the material and geometrical parameters of the diffractive structure one can control the contrast transmission that characterizes the degree of the asymmetry. This property of the structure is demonstrated by the results of computer simulation calculations.

\section{Theoretical model}
\label{sec:model}

\begin{figure}[h]
\epsfig{file=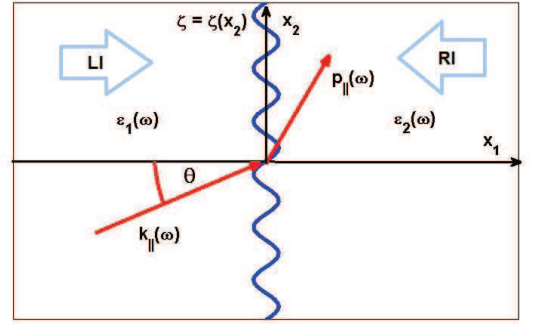,angle=0,width=\columnwidth}
\caption{(Color online) Surface structure studied in this paper.}
\label{fig1}
\vspace{0.1cm}
\end{figure}

The system we consider in this paper consists of vacuum in the region $x_3 >0$, while the region $x_3 <0$ consists of two metalllic segments characterized by frequency-dependent dielectric functions $\epsilon_1(\omega)$ and $\epsilon_2(\omega)$. The interface between the two metals is characterized by the profile function $\zeta(x_2)$ which separates the neighboring metals in the regions $x_1 < \zeta(x_2)$  and $x_1 > \zeta(x_2)$ - see  Fig.\ref{fig1}

We denote the third component of the electric field of a surface plasmon polariton in the vacuum region $x_3 > 0$ evaluated on the surface $x_3 = 0$, $E^{>}_{3}(x_1,x_2,0|\omega)$, by $E^{>}_{3}(x_1,x_2,|\omega)$. The equation it satisfies in the region $x_1 < \zeta(x_2)$ is

\begin{eqnarray}
\left(\frac{\partial^2}{\partial x^2_1} + \frac{\partial^2}{\partial x^2_2} + k^2_{\|}(\omega)\right)E^{>}_3(x_1,x_2|\omega) = 0,
\label{Eq1}
\end{eqnarray}
where $k_{\|}(\omega) = \frac{\omega}{c}\left[1 - \frac{1}{\epsilon_1(\omega)}\right]^{\frac{1}{2}}.$

The incident field is a solution of this equation, and we write it in the form



\begin{eqnarray}
E^{>}_{3}(x_1,x_2,|\omega)_{inc} = -\frac{c}{\omega}k_{\|}e^{[i\alpha_1(k_2,\omega)x_1 + ik_2x_2)]} .
\label{Eq2}
\end{eqnarray}

where

\begin{subequations}
\begin{eqnarray}
k_1 = \alpha_1(k_2,\omega) = [k^2_{\|}(\omega) - k_2^2]^{\frac{1}{2}} \ \ \ \ k_2^2 < k^2_{\|}(\omega)
\\
 = i[k_2^2 - k^2_{\|}(\omega)]^{\frac{1}{2}} \ \ \ \ k_2^2 > k^2_{\|}(\omega).
\label{Eq3}
\end{eqnarray}
\end{subequations}


The equation $E^{>}_{3}(x_1,x_2,|\omega)$ satisfies in the region $x_1 > \zeta(x_2) , x_3 = 0$ is

\begin{eqnarray}
\left(\frac{\partial^2}{\partial x^2_1} + \frac{\partial^2}{\partial x^2_2} + p^2_{\|}(\omega)\right)E^{>}_3(x_1,x_2|\omega) = 0,
\label{Eq4}
\end{eqnarray}
where $p_{\|}(\omega) = \frac{\omega}{c}\left[1 - \frac{1}{\epsilon_2(\omega)}\right]^{\frac{1}{2}}.$


The transmitted field is the solution of this equation, which we write as


\begin{eqnarray}
E^{>}_{3}(x_1,x_2,|\omega)_{tr} = \int^{\infty}_{-\infty}{\frac{dq_2}{2\pi}T(q_2)e^{i[p_{\|}^2(\omega)-q^2_2]^\frac{1}{2}x_1 + iq_2x_2}}.
\label{Eq5}
\end{eqnarray}


Now

\begin{eqnarray}
E^{>}_{3}(0^+,x_2,|\omega)_{tr} = \int^{\infty}_{-\infty}{\frac{dq_2}{2\pi}T(q_2)e^{iq_2x_2}},
\label{Eq6}
\end{eqnarray}

so that

\begin{eqnarray}
T(q_2) = \int^{\infty}_{-\infty}{dx'_2e^{-iq_2x'_2}E^{>}_{3}(0^+,x'_2,|\omega)_{tr}}.
\label{Eq7}
\end{eqnarray}

Therefore
\small
\begin{align}
E^{>}_{3}(x_1,x_2,|\omega)_{tr} & = \int^{\infty}_{-\infty}dx'_2\left\{
\int^{\infty}_{-\infty}{\frac{dq_2}{2\pi}e^{i[p^2_{\|}(\omega)-q^2_2]^\frac{1}{2}x_1 + iq_2(x_2-x^{'}_2)}}\right\}  \nonumber
\\
& \times E^{>}_{3}(0^+,x'_2,|\omega)_{tr}.
\label{Eq8}
\end{align}
\normalsize
The thin phase screen model~\cite{thin_phase}  states that
\small
\begin{eqnarray}
E^{>}_{3}(0^+,x_2,|\omega)_{tr} = e^{i\Delta n \frac{\omega}{c}\zeta(x_2)}E^{>}_{3}(0^-,x_2,|\omega)_{inc} ,
\label{Eq9}
\end{eqnarray}
\normalsize
where

\begin{align}
    \Delta n = n_1(\omega) - n_2(\omega) = \left[1 - \frac{1}{[\epsilon_1(\omega}\right]^{\frac{1}{2}} - \left[1 - \frac{1}{\epsilon_2(\omega)}\right]^{\frac{1}{2}}.
\label{Eq10}
\end{align}

By inserting the relation given by Eq.(\ref{Eq9}) into Eq.(\ref{Eq8}) one obtains for the third component of the electric field of the transmitted SPP in the thin phase screen model

\begin{align}
E^{>}_{3}(x_1,x_2,|\omega)_{tr} & = -\frac{c}{\omega}k_{\|}\int^{\infty}_{-\infty}{\frac{dq_2}{2\pi}e^{i[p^2_{\|}-q^2_2]^\frac{1}{2}x_1 + iq_2x_2}}  \nonumber \\
& \times \int^{\infty}_{-\infty}dx'_2e^{i(q_2-k_2)x^{'}_2}e^{i\Delta n \frac{\omega}{c}\zeta(x^{'}_2)}.
\label{Eq11}
\end{align}

We assume the surface profile function $\zeta(x_2)$ is a periodic function of $x_2$, $\zeta(x_2 + a) = \zeta(x_2)$ , where $a$ is the period. Then the second integral on the right-hand side(RHS) of Eq.(\ref{Eq11}) can be replaced by the following sum
\begin{align}
\int^{\infty}_{-\infty}dx_2e^{-i(q_2-k_2)x_2}e^{i\Delta n \frac{\omega}{c}\zeta(x_2)}  \hspace{4cm}\nonumber \\
= \sum^{\infty}_{n=-\infty}\int^{\left(n+\frac{1}{2}\right)a}_{\left(n-\frac{1}{2}\right)a} e^{-i(q_2-k_2)x_2 + i\Delta n \frac{\omega}{c}\zeta(x_2)}. \hspace{2cm}\nonumber
\end{align}
Then by introducing  $x_2 = na + x$, the RHS of this equation becomes
\small
\begin{align}
\sum^{\infty}_{m=-\infty}{2\pi \delta \left(q_2 - k_2 - \frac{2\pi m}{a}\right)}
\frac{1}{a}\int^{\frac{a}{2}}_{-\frac{a}{2}}dxe^{-i\frac{2\pi m x}{a}}e^{i\Delta n \frac{\omega}{c}\zeta(x)}.
\label{Eq12}
\end{align}
\normalsize
Therefore

\begin{align}
E^{>}_{3}(x_1,x_2,|\omega)_{tr} & = -\frac{c}{\omega}k_{\|}\int^{\infty}_{-\infty}{\frac{dq_2}{2\pi}e^{i[p^2_{\|}-q^2_2]^\frac{1}{2}x_1 + iq_2x_2}}  \nonumber
\end{align}
\begin{align}
\times \sum^{\infty}_{m=-\infty}{2\pi \delta (q_2 - k_{2m})} \frac{1}{a}\int^{\frac{a}{2}}_{-\frac{a}{2}}dxe^{-i\frac{2\pi m x}{a}}e^{i\Delta n \frac{c}{\omega}\zeta(x)} \nonumber
\end{align}
\begin{align}
= -\frac{\omega}{c}k_{\|} \sum^{\infty}_{m=-\infty}e^{i[p^2_{\|}-k^2_{2m}]^\frac{1}{2}x_1 + ik_{2m}x_2}  \nonumber  \\
\times \frac{1}{a}\int^{\frac{a}{2}}_{-\frac{a}{2}}dxe^{-i\frac{2\pi m x}{a}}e^{i\Delta n \frac{\omega}{c}\zeta(x)},
\label{Eq13}
\end{align}
where $k_{2m} = k_2 + \frac{2\pi m}{a}$. If $\zeta(x)$ is an even function of $x_1$, $\zeta(-x_1) = \zeta(x_1)$,

\begin{align}
E^{>}_{3}(x_1,x_2,|\omega)_{tr} = -\frac{\omega}{c}k_{\|} \sum^{\infty}_{m=-\infty}e^{i[p^2_{\|}-k^2_{2m}]^\frac{1}{2}x_1 + ik_{2m}x_2}  \nonumber \\
\times \frac{2}{a}\int^{\frac{a}{2}}_{0}dx\cos\frac{2\pi m x}{a}e^{i\Delta n \frac{\omega}{c}\zeta(x)}.
\label{Eq14}
\end{align}

In the case of a periodic rectangular profile,

\begin{align}
\zeta = & \zeta_0 & 0 < x < \frac{a}{4} \nonumber \\
\zeta = & -\zeta_0 & \frac{a}{4} < x < \frac{a}{2},
\label{Eq15}
\end{align}
we obtain for the zero-order and the higher order terms in the expansion given by Eqs.(\ref{Eq13}) and (\ref{Eq14})

\begin{align}
 \frac{1}{a}\int^{\frac{a}{2}}_{-\frac{a}{2}}dxe^{-i\frac{2\pi m x}{a}}e^{i\Delta n(\omega) \frac{\omega}{c}\zeta(x)}  \hspace{6cm}\nonumber \\
 = \cos \alpha \zeta_0 \hspace{1.5cm}  m = 0 \hspace{6cm}\nonumber \\
 = \frac{2i}{\pi}(-1)^k \frac{\sin \alpha \zeta_0}{2k+1} \hspace{0.5cm} m = 2k+1 \hspace{0.5cm} k = 0,\pm 1,\pm2,\ldots \hspace{2cm}.
\label{Eq16}
\end{align}
where $\alpha = \Delta n(\omega) \frac{\omega}{c}$. Therefore the amplitude of the zero-order term given by Eq. (\ref{Eq16}) vanishes when the depth of the rectangular grating
$\zeta_0 = \pi/(2\alpha)$, which states that the critical value of the modulation depth $\zeta_0$ at given wavelength is inversely proportional to the refractive index contrast $\Delta n$

\begin{eqnarray}
\zeta_0 = \frac{\lambda}{4\Delta n}.
\label{Eq17}
\end{eqnarray}

The amplitudes of the higher-order Bragg beams accordingly follow the sine-like behavior of Eq. (16) and thus, for example first-order waves have a non-zero amplitude at the same wavelength at which the zero-order beam vanishes.


\section{Results: thin phase screen model}
\label{results}

 We demonstrate asymmetric transmission characteristics of a SPP by the use of the thin phase screen model in the case of the periodically modulated interface characterized by the rectangular periodic profile given by Eq.(\ref{Eq15}). A key idea underlying asymmetric transmission in such a planar grating structure is the suppression of the zero-order Bragg mode in a certain frequency range, which occurs at a given wavelength for the critical value of the modulation depth $\zeta_0$ given by  Eq.(\ref{Eq17}). Then one can modify the diffraction efficiencies of the first-order Bragg beams by varying the period $a$ of the grating and/or the angle of incidence $\theta$.

\subsection{Normal incidence}
\label{thin phase_normal}

\begin{figure}[h]
\epsfig{file=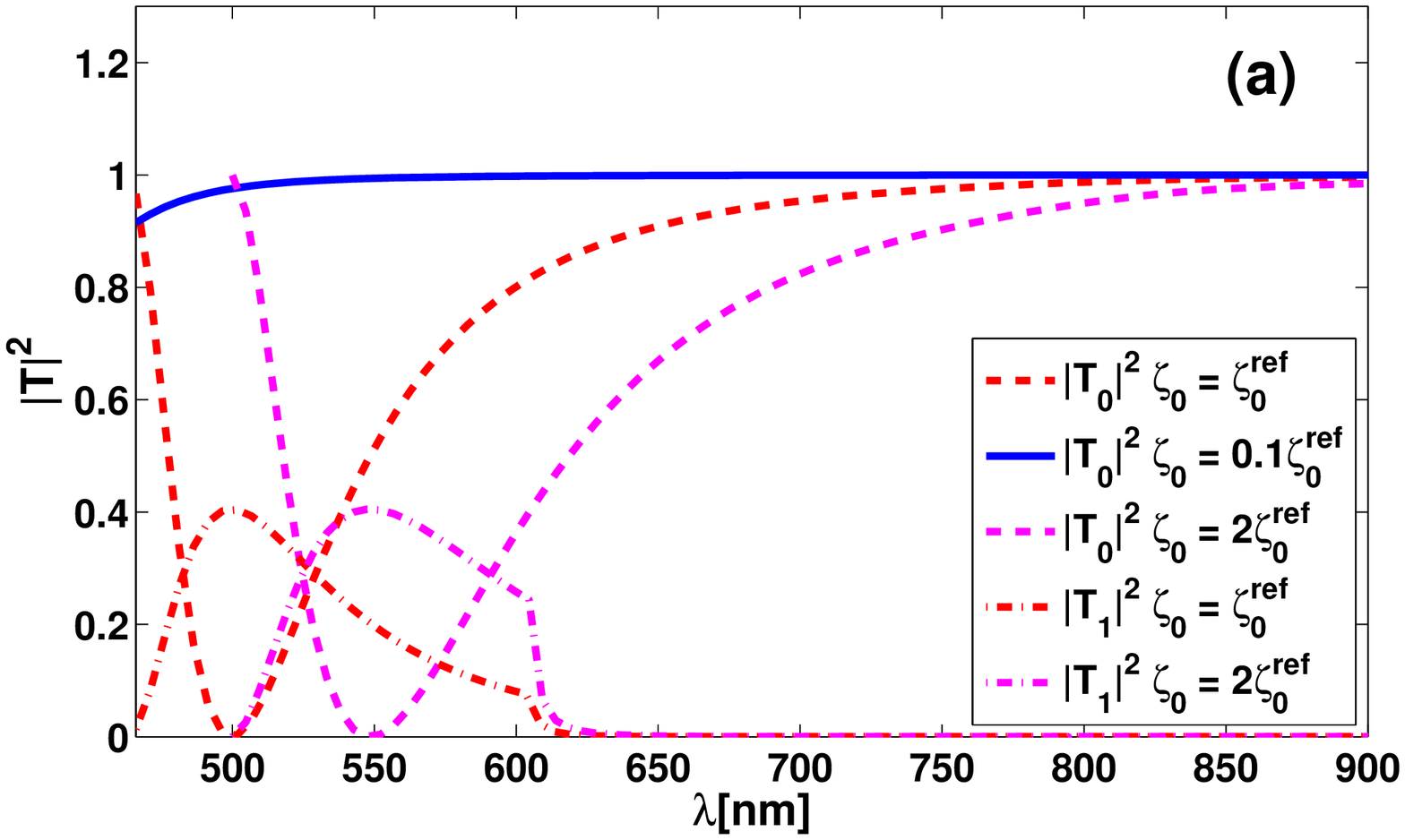,angle=0,width=\columnwidth}
\epsfig{file=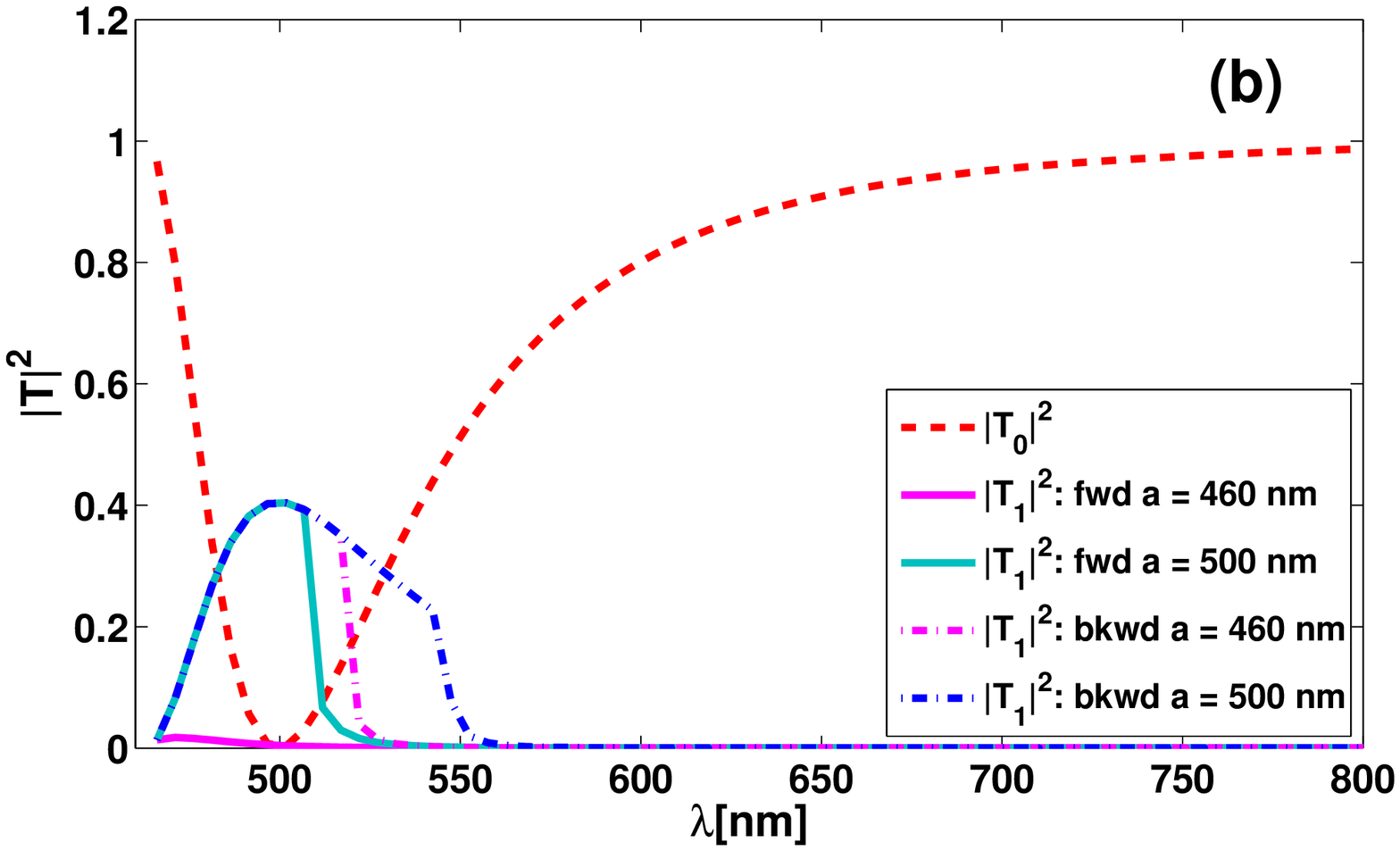,angle=0,width=\columnwidth}
\epsfig{file=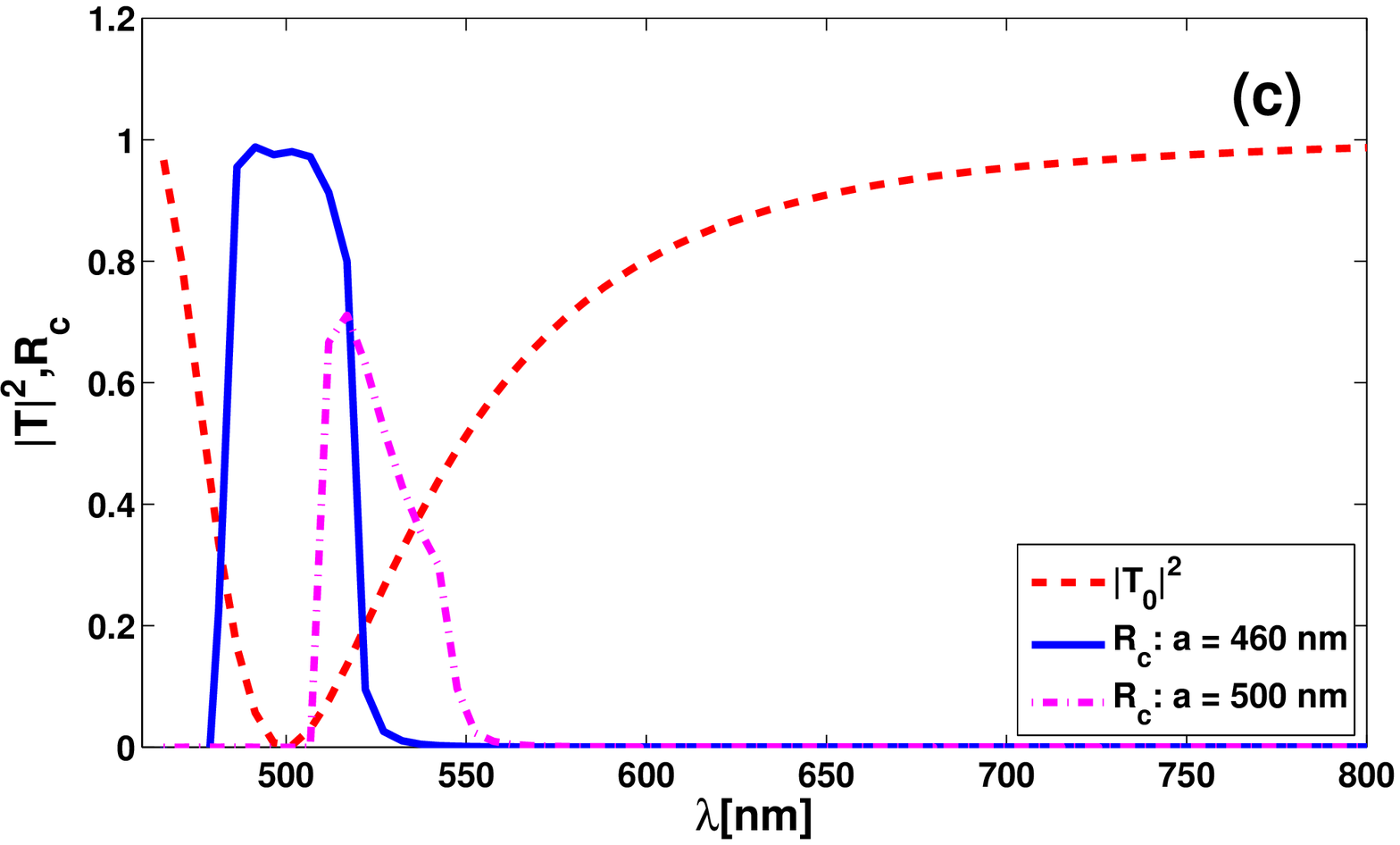,angle=0,width=\columnwidth}
\caption{(Color online) (a) Transmittances of the zero- and first-order Bragg beams vs. wavelength $\lambda$ obtained from the thin phase screen model as functions of the modulation depth $\zeta_0$ when $a = 600 \ nm$; (b) the transmittances of both forward- and backward propagating first-order Bragg beams vs. $\lambda$  as functions of the period $a$; (c) transmittance contrast ratio $R_c$ between the forward- and backward-propagating first-order Bragg beams as functions of the period $a$ when the latter varies in the frequency range  $460 \ nm < a < 500 \ nm$. The red dashed line in (b) and (c) indicates the transmittance of the zero-order Bragg beam for $\zeta_0  = 887 \ nm$ and $a = 600 \ nm$.}
\label{fig2_new}
\vspace{0.1cm}
\end{figure}

In the following we consider the case of normal incidence and study transmission through a lamellar grating at an Au/Al interface with a period $a = 600 \ nm$. By using the refractive index contrast $\Delta n = 0.140906$ between Au and Al for the wavelength $\lambda = 500 \ nm$, one obtains the critical modulation depth $\zeta_0^{ref} = 887 \ nm$ at which the amplitude of the transmitted zero-order Bragg beam vanishes according to relation (\ref{Eq17}). The behavior of the zero-order Bragg beam is shown in Fig.\ref{fig2_new}(a) where the transmittance $|T_0|^2$(red dashed curve) reveals a profound suppression in the vicinity of the reference wavelength $\lambda_{ref} = 500 \ nm$. This feature reflects the strong dependence of $|T_0|^2$ on the modulation depth $\zeta_0$, which is evident in comparison with results obtained for the interface characterized by a 10$\times$ smaller modulation depth $\zeta_0 = 0.1\zeta_0^{ref}$ for which the interface becomes transparent in the wavelength range considered(full blue curve).
The frequency range in which the suppression of the 0-order mode occurs can be modified by varying the modulation depth $\zeta_0$ as is demonstrated in Fig.~\ref{fig2_new}(a) where the minimum in the transmittance $|T_0|^2$ associated with the zero-order SPP Bragg beam is red-shifted to the wavelength $\lambda = 550 \ nm$ when the modulation depth is increased by a factor of 2 - see the magenta dashed curve in Fig.~\ref{fig2_new}(a). The dependence of the critical depth $\zeta_0$ on the wavelength $\lambda$ is affected by the strongly dispersive behavior of the dielectric functions of both metals in the wavelength range considered, which give rise to a decreasing refractive index contrast as the wavelength is increased. In fact, the choice of the Au/Al interface stems from the fact that these two metals yield a large refractive index contrast $\Delta n$ at the wavelength $\lambda = 500 \ nm$. For example, the transmittance of the zero-order Bragg beam through a Au/Ag interface to a large extent resembles that for a Au/Al interface - see Fig.~\ref{fig2_new}(a).  However, it requires at the same wavelength a significantly larger modulation depth $\zeta_0 = 1350 \ nm$, which renders this configuration technologically more challenging from the point of view of both its numerical and experimental verification.

We offer the following simple intuitive explanation of this result.  The difference in optical path lengths between the waves incident on the regions of the surface where $\zeta(x_2) = -\zeta_0$  and where $\zeta(x_2) = +\zeta_0$  is equal to $2\zeta_0 \Delta n$.  When $\zeta_0$    has its critical value $\zeta_0 = \lambda/(4 \Delta n)$  this difference in path lengths equals $\lambda/2$ .  The resulting destructive interference of these two waves leads to the suppression of the zero-order beam.  An analogous interpretation can be given for the higher-order beams.  For example, the first-order waves incident on the regions of the surface where $\zeta(x_2) = -\zeta_0$  and where $\zeta(x_2) = \zeta_0$   are in phase, and thus interfere constructively at the same wavelength at which the zero-order beam vanishes.

Now we inspect the possibility of achieving asymmetry in the transmittance of the first-order Bragg beam in the case of normal incidence. We found that a lamellar grating at an Au/Al interface with the period $a = 600 \ nm$ and modulation depth $\zeta_0 = 887 \ nm$ exhibits a suppression of the zero-order Bragg beam in a certain frequency range, while the transmittance of the 1-order Bragg beams becomes significantly enhanced in this frequency range in comparison with that associated with a shallow modulation $\zeta_0 = 0.1\zeta_0^{ref}$ -- see the dashed-lines in Fig.~\ref{fig2_new}(b), which correspond to the transmittances of the first-order Bragg beams $|T_{\pm 1}|^2$ when the modulation depth $\zeta_0 = \zeta_0^{ref}$(red dash-dotted line) and  $\zeta_0 = 2\zeta_0^{ref}$(magenta dash-dotted line). The transmittances $|T_{\pm 1}|^2$ for $a = 600 \ nm$ are nearly identical for the first-order Bragg beams propagating in the opposite directions. However, they become significantly different when the period $a$ is decreased. Specifically, the transmittances $|T_{\pm 1}|^2$ associated with the first-order Bragg beams  propagating in the $Au \rightarrow Al$ direction(LI) vanish when the period $a = 460 \ nm$(solid magenta line in Fig.~\ref{fig2_new}(b), and gradually recover when the period $a$ is increased -- see the transmittances $|T_{\pm 1}|^2$ for $a = 480 \ nm$ and $a = 500 \ nm$ indicated by the solid blue and green lines in Fig.~\ref{fig2_new}(b), respectively. The transmittances $|T_{\pm 1}|^2$ for the backward-propagating first-order Bragg beams along the $Al \rightarrow Au$ direction(RI) exhibit a similar dependence on the magnitude of the period $a$, except that the upper wavelength cutoff at which the transmittance vanishes is red-shifted -- see the transmittances $|T_{\pm 1}|^2$ for $a = 460 \ nm$, $a = 480 \ nm$ and $a = 500 \ nm$ indicated by dash-dotted magenta, blue and green lines in Fig.~\ref{fig2_new}(b).
\begin{figure}[h]
\epsfig{file=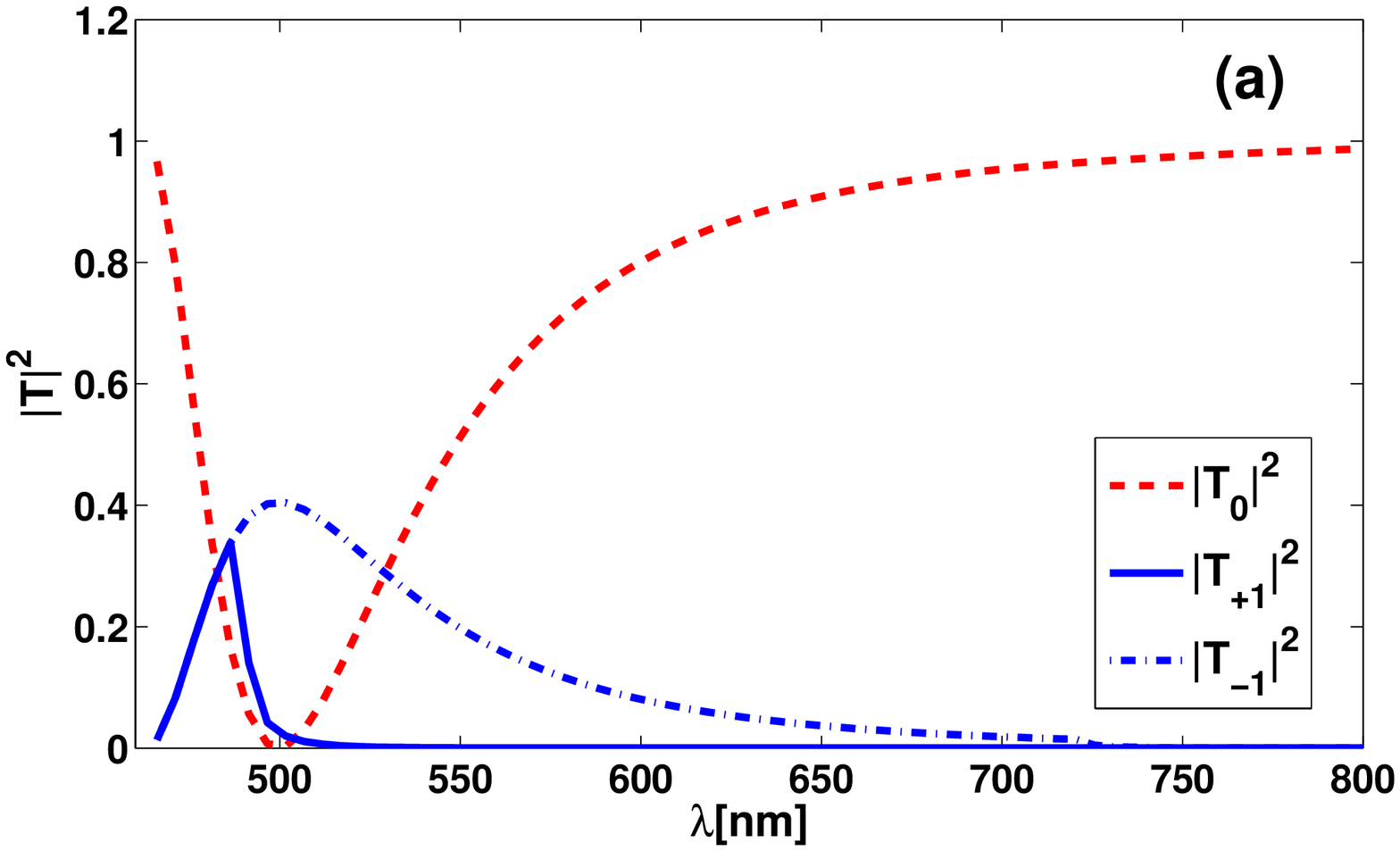,angle=0,width=\columnwidth}
\epsfig{file=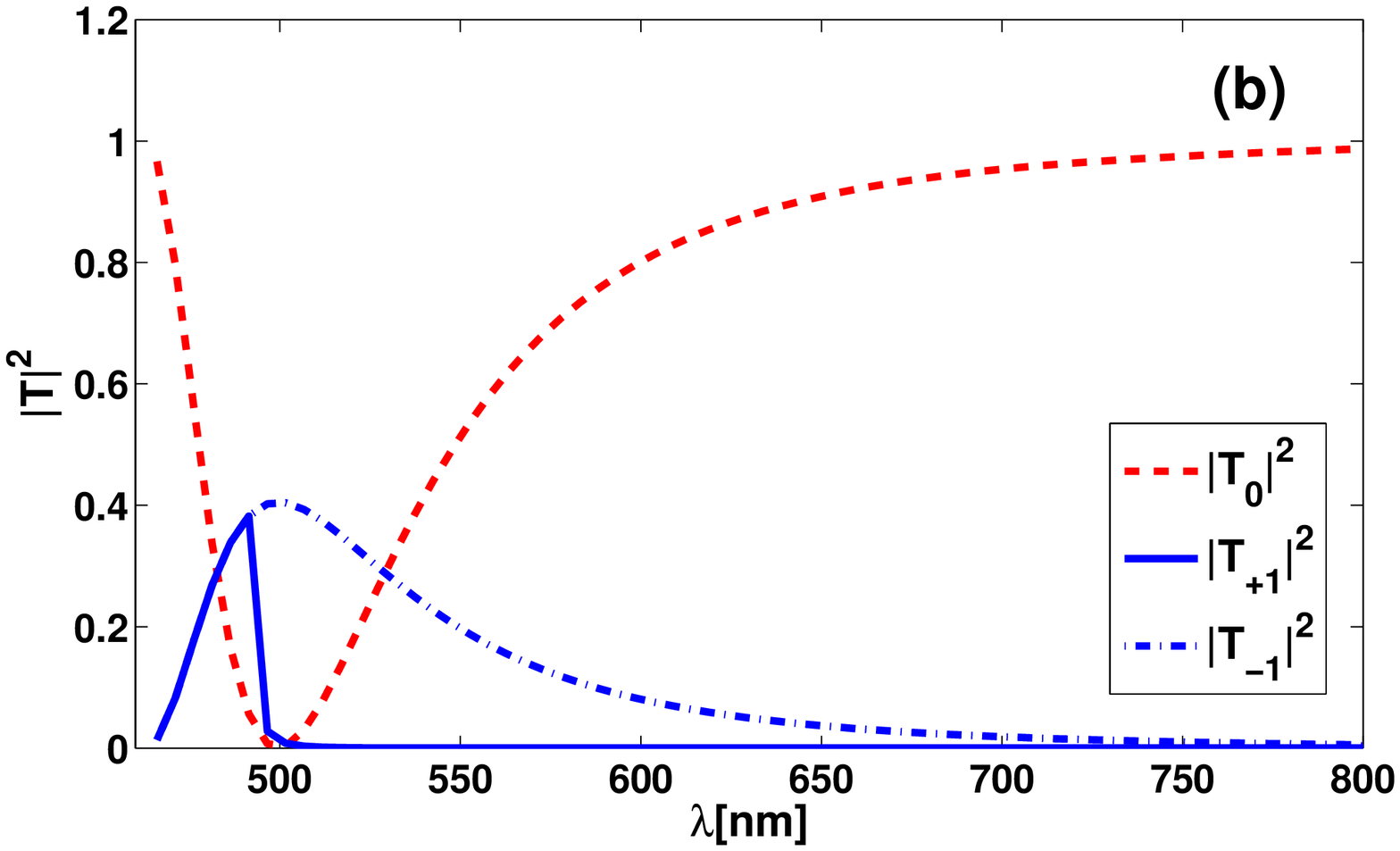,angle=0,width=\columnwidth}
\epsfig{file=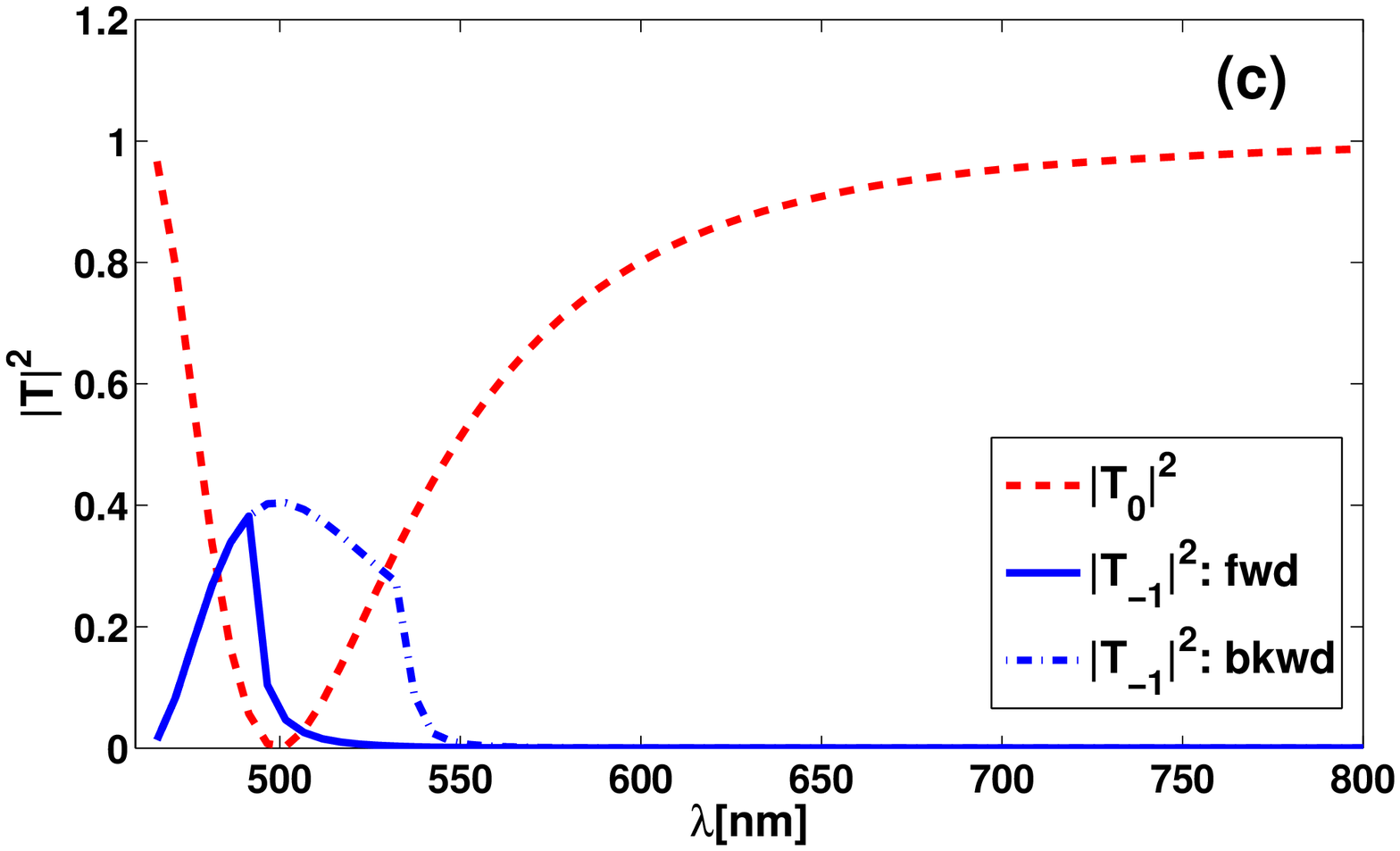,angle=0,width=\columnwidth}
\caption{(Color online) (a) Transmittances of the $+1$ and $-1$ forward Bragg beams vs. wavelength $\lambda$ incident on the Au/Al lamellar grating at angle $\theta = 22.5^{o}$ from the left(solid blue line) and from the right side(dash-dotted blue line) obtained from the thin phase screen model; (b) the same except the transmittances correspond to the backward propagating $+1$ and $-1$ Bragg beams at the angle $\theta = 35^{o}$; (c)the transmittances of both the forward(solid blue line) and backward(dash-dotted blue line) propagating $-1$ Bragg beams vs. $\lambda$  when the lattice constant $a = 430 \ nm$. The red dashed line in all panels indicates the transmittance of the zero-order Bragg beam for $\zeta_0  = 887 \ nm$ and $a = 430 \ nm$.}
\label{fig3}
\vspace{0.1cm}
\end{figure}
We describe the asymmetry in the transmittance of the first-order Bragg beams in terms of the transmissivity contrast  ratio defined as
\begin{eqnarray}
R_c = \frac{T_L - T_R}{T_L + T_R},
\label{Eq18}
\end{eqnarray}
where $T_L$ and $T_R$ denote the transmittances of the forward and backward-propagating first-order Bragg beams. The mismatch in the transmittances $|T_{\pm}1|^2$ demonstrated in Fig.~\ref{fig2_new}(b) implies a significant transmissivity contrast $R_c$ between the counter-propagating beams in the wavelength range $490 \ nm < \lambda < 550 \ nm$ -- see  Fig.~\ref{fig2_new}(c). We note that the size of the asymmetry and wavelength range in which this effect occurs depend on the period $a$, namely when the latter is decreased (increased) the size of the asymmetry decreases and the corresponding wavelength range is shifted towards smaller (larger) wavelengths. This is demonstrated in Fig.~\ref{fig2_new}(c), where we depict the transmissivity contrast $R_c$  for the periods $a = 460 \ nm$, $a = 480 \ nm$, and $a = 500 \ nm$ indicated by the magenta solid lines, and the dashed blue and green lines, respectively.  We also note that the degree of asymmetry decreases as the wavelength increases.

\subsection{Oblique incidence}
\label{thin phase_oblique}

In this section we explore regimes which support asymmetry in the transmittance of the first-order Bragg beams in the case of oblique incidence. We consider our reference system to be a lamellar grating at an Au/Al interface with the period $a = 600 \ nm$ and modulation depth $\zeta_0 = 887 \ nm$. It is well known that at normal incidence the transmittances associated with the $+1$ and $-1$ Bragg beams are identical, and the two waves propagate along directions that are symmetric with respect to the normal to the interface. When the angle of incidence is changed the transmittances exhibit different dependencies on the wavelength, as is illustrated in Figs.~\ref{fig3}(a) and (b) for the forward(LI) and backward(RI)-propagating waves, respectively. Namely, for the angle of incidence $\theta = 22.5^{o}$, the transmittance associated with the $+1$ Bragg beam(solid blue line in Fig.~\ref{fig3}(a)) reveals a cutoff at the wavelength $\lambda = 500 \ nm$, while that belonging to the $-1$ Bragg beam remains unaffected in the frequency range where the zero-order beam is suppressed and shows a slowly decaying tail as the wavelength is increased[dash-dotted blue line in Fig.~\ref{fig3}(a)]. We observe similar behavior in the case of the backward propagating wave(RI) -- see Fig.~\ref{fig3}(b) -- except that the equivalent wavelength cutoff for the transmittance associated with the $+1$ Bragg beam is achieved for a larger angle of incidence $\theta = 35^{o}$ [solid blue curve in Fig.~\ref{fig3}(b)] while the transmittance associated with the $-1$ Bragg beam[dash-dotted blue curve in Fig.~\ref{fig3}(b)] resembles the behavior of the corresponding counter-propagating wave.

The results presented in Figs.~\ref{fig3}(a) and (b) show that the $+1$ SPP Bragg beams incident in a certain range of the angle of incidence become evanescent, and only the $-1$ Bragg beams are propagating in this regime. Referring to the results shown in Figs.~\ref{fig2_new}(b),(c) it is obvious that one can achieve an asymmetry in the transmittance of the $-1$ Bragg beams by varying the size of the period $a$ as is illustrated in Fig.~\ref{fig3}(c), where the transmittance associated with the forward propagating $-1$ Bragg beam[solid blue curve in Fig.~\ref{fig3}(c)] indicates that the wave becomes evanescent for $\lambda > 500 \ nm$ when the period $a = 430 \ nm$,  while the $-1$ SPP Bragg beam incident from the opposite side remains propagating[dash-dotted blue curve in Fig.~\ref{fig3}(c)] in the wavelength range where both the zero-order and $+1$ Bragg beams are suppressed.
\begin{figure}[h]
\epsfig{file=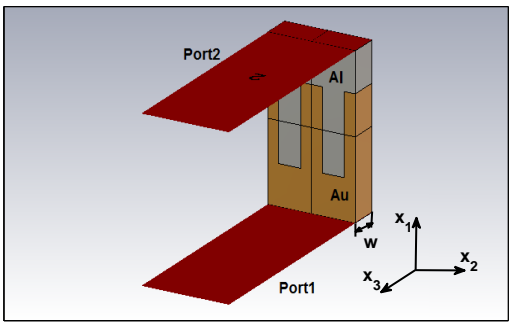,angle=0,width=\columnwidth}
\caption{(Color online) Computational domain used for numerical simulations of the surface grating structures. Port1 and Port2 denote the waveguide ports used in the frequency-domain solver.}
\label{fig4_cst}
\vspace{0.1cm}
\end{figure}

\section{Results: Finite-element frequency-domain method}
\label{CST}

In order to verify the predictions of the thin phase screen model we calculate numerically the transmittance of the SPP propagating across the modulated Au/Al interface shown in Fig.\ref{fig1}. We employ the CST frequency domain solver~\cite{CST}.
The interface between a semi-infinite vacuum and a semi-infinite metal depicted in Fig.\ref{fig1} is replaced by the computational domain shown in Fig.\ref{fig4_cst}, where a finite thickness of the metal region $w$ is assumed, typically $w = 400 \ nm$, and periodic boundary conditions are applied along the $x_2$ axis  .
We assume in this configuration a surface plasmon polariton propagating in the forward direction along the $x_1$ axis, incident on the structure from the bottom, which corresponds to the left incidence(LI) indicated in Fig.\ref{fig1}, while a surface plasmon polariton propagating in the backward direction along the $x_1$ axis is incident on the structure from the top, which corresponds to the right incidence(RI).
\begin{figure}[h]
\epsfig{file=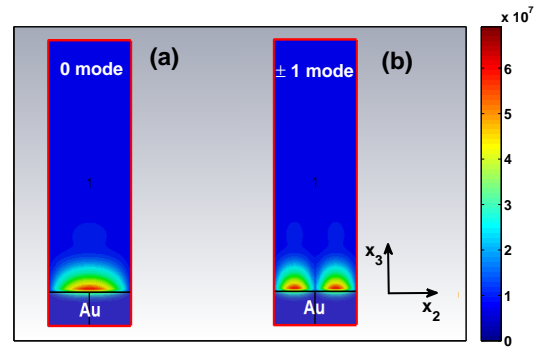,angle=0,width=\columnwidth}
\caption{(Color online) Spatial intensity distribution of the electric fields associated with the (a) zero-order and (b) $\pm 1$  SPP Bragg modes belonging to the shielded Au waveguide port obtained from the frequency-domain solver. The SPP modes with the wavelength $\lambda = 500 \,nm$ incident on the structure depicted in Fig.\ref{fig4_cst} from the bottom correspond to the left incidence(LI) indicated in Fig.\ref{fig1}. The electric fields are in $Vm^{-1}$ units.}
\label{fig4_Au}
\vspace{0.1cm}
\end{figure}
We impose the electric wall boundary condition $E_t = 0$ along the $x_1^+$, $x_1^-$ and  $x_3$-axes and open boundary conditions along the $x_3^+$ axis in the region $x_3 > w$. As the excitation source we choose waveguide ports attached to the bottom(Au)  and the top(Al) of the metallic region. The waveguide ports represent a special kind of boundary condition of the calculation domain, which requires enclosing the entire domain filled with the electric field. This kind of port simulates an infinitely long waveguide connected to the structure. Then the eigenmode solver allows calculating the exact port modes within these boundaries, including the wave numbers(propagation constants) of these modes $k_1^m = \frac{\omega}{c} n_{eff}^m$, or more precisely their effective indices $n^m_{eff}$.
\begin{figure}[h]
\epsfig{file=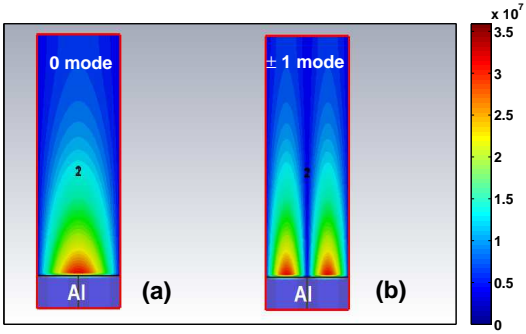,angle=0,width=\columnwidth}
\caption{(Color online) Spatial intensity distributions of the electric fields associated with the (a) zero-order and (b) $\pm 1$ SPP Bragg modes belonging to the Al waveguide port obtained from the frequency-domain solver. The SPP modes with the wavelength $\lambda = 500 \,nm$ incident on the structure depicted in Fig.\ref{fig4_cst} from the top correspond to the right incidence(RI) indicated in Fig.\ref{fig1}. The electric fields are in $Vm^{-1}$ units.}
\label{fig4_Al}
\vspace{0.1cm}
\end{figure}
In calculating the transmittance of the structure we take into account a sufficiently large number of modes which form a finite subset of the eigenmodes supported by the structure. We note that the majority of the modes are radiative, which together with ohmic losses represent two channels of decay that determine the lifetime of the SPP. We focus on the surface modes with electric fields strongly confined to the surface $x_1x_2$, which correspond to the lowest order diffraction orders of the SPP propagating along the metal-vacuum interface.

\subsection{Normal incidence}
\label{CST_normal}

\begin{figure}[h]
\epsfig{file=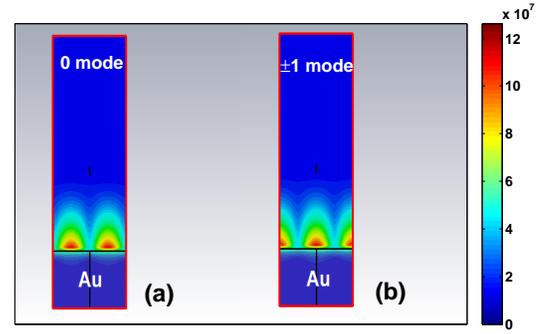,angle=0,width=\columnwidth}
\caption{(Color online) Spatial intensity distributions of the electric fields associated with the (a) -1 and (b) $+1$ SPP Bragg modes belonging to the Au waveguide port obtained from the frequency-domain solver in the regime of unshielded waveguide ports. Both nearly degenerate modes correspond to the doubly degenerate $\pm 1$ modes displayed in Fig.\ref{fig4_Au}. The SPP modes with the wavelength $\lambda = 500 \,nm$ are incident on the structure depicted in Fig.\ref{fig4_cst} from the bottom, which corresponds to the left incidence(LI) shown in Fig.\ref{fig1}.The electric fields are in $Vm^{-1}$ units.}
\label{fig5}
\vspace{0.1cm}
\end{figure}

 In the following we consider the case of normal incidence. In Figs.\ref{fig4_Au} and \ref{fig4_Al} we show the intensities of the electric field associated with the zero and $\pm 1$ SPP modes belonging to the Au and Al waveguide ports, respectively. One can see that the lowest order SPP modes associated with the Au port are more confined to the surface than those belonging to the opposite Al port. The field intensity patterns shown in Fig.\ref{fig4_Au}(b) and \ref{fig4_Al}(b)  correspond to the doubly degenerate $\pm 1$ modes obtained from the frequency-domain solver in the case of normal incidence with shielded ports. The latter feature represents a special kind of boundary condition, which restricts the influence of the higher order radiative modes but at the same time does not allow calculating the eigenmodes for oblique incidence when periodic boundary conditions along the $x_1$ axis are imposed. In the unshielded regime, the asymmetry of the structure leads to the splitting of the doubly degenerate $\pm 1$ Bragg modes, and the frequency-domain solver yields singlets corresponding to the $+1$, $-1$ waveguide eigenmodes. The nearly degenerate modes belonging to the Au port possess the effective indices $n_{eff}^f(\pm 1)_{\theta = 0^{\circ}} = 0.39$ -- see Fig.\ref{fig5} -- and the degenerate $+1$ and $-1$ modes belonging to the Al port possess the effective indices $n_{eff}^b(\pm 1)_{\theta= 0^{\circ}} = 0.38$. Here the subscripts $f$ and $b$ denote forward and backward propagating modes belonging to the Au and Al ports, respectively.

To understand the propagation of the SPP through the grating region represented by the modulated Au/Al interface with a rectangular profile it is helpful to employ the results of the modal method described in detail in Ref.~\onlinecite{Sheng}. In this method the incident, transmitted and reflected fields are expressed in terms of the eigenmodes of the structure and thus cannot be described solely by the well-known grating equation, since for the calculation of the amplitudes of the diffraction orders the EM field inside the grating region has to be considered. According to this approach, the propagation of the wave in the $x_1$ direction through the grating region resembles that of a simple slab waveguide that is able to guide a discrete set of modes. The similarity between the field distributions associated with the incident and excited modes is given by an overlap integral,  while  matching of the effective indices is characterized by the difference between the wave number of the incident wave $k_1^{in} = \frac{\omega}{c} \cos \phi_{in}$ and the wave number $k_1^m$, where $\phi_{in}$ is angle of incidence.  Both factors determine how much energy of the incident wave is coupled to a specific mode.

%
%

The transmittance of the structure shown in Fig.\ref{fig4_cst} can be described in terms of diffraction efficiencies calculated as the intensities of the transmitted higher-order Bragg beams divided by the intensity of the incident wave. The frequency-domain  solver calculates the scattering matrix between the two sets of eigenmodes associated with both waveguide ports.
We first implemented our reference structure with a lamellar grating at an Au/Al interface with the period $a = 600 \ nm$ and modulation depth $\zeta_0 = 887 \ nm$ which, according to the results of the thin phase screen model, does not support propagation of the zero-order Bragg beam in a certain frequency range. We confirmed this by evaluating the transmittance associated with the eigemnodes that correspond to the zero-order SPP mode supported by both the Au and Al ports, and showed that they possess vanishing values $T_0 \sim 0.01$.
\begin{figure}[h]
\epsfig{file=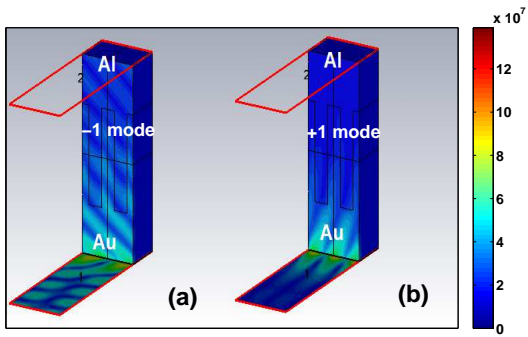,angle=0,width=\columnwidth}
\caption{(Color online) Spatial intensity distributions of the electric fields along the waveguide structure associated with (a) the $-1$ and (b) the $+1$ forward propagating SPP Bragg modes incident at an angle $\theta = 75^{\circ}$. The electric fields are in $Vm^{-1}$ units.}
\label{fig7}
\vspace{0.1cm}
\end{figure}
%



\subsection{Oblique incidence}
\label{CST_oblique}

In this subsection we consider the case of oblique incidence, namely we study how the diffraction efficiencies depend on the angle of incidence $\theta$. We focus on the $+1$ and $-1$ modes, which reveal strong and profoundly different dependencies on the incidence angle and simultaneously reveal a dependence on the direction of propagation. We first consider the modes belonging to the Au port i.e. the forward propagating waves.  The transmittance of the  $+1$ mode propagating in the forward direction $Au \rightarrow Al$ in the case of normal incidence is somewhat larger $T_f(+1)_{\theta = 0^{\circ}} = 0.26$ than that of the $-1$ mode $T_f(-1)_{\theta = 0^{\circ}} = 0.19$.
When the angle of the incidence is increased, the difference in the transmittances of the $+1$ and $-1$ modes increases, and for sufficiently large angles of incidence $\theta \gtrsim 60^{\circ}$ the transmittance of the $+1$ mode vanishes.  Specifically, the transmittance belonging to the $-1$ mode is slightly increased  $T_f(-1)_{\theta = 75^{\circ}} = 0.28$ when $\theta = 75^{\circ}$, while the transmittance of the $+1$ mode becomes negligible $T_f(+1)_{\theta = 75^{\circ}} = 0.0001$.
This behavior is demonstrated in Fig.\ref{fig7}, where the spatial distributions of the electric field intensities along the waveguide structure for the $-1$ (a) and $+1$ (b) Bragg beams are shown for the Au waveguide port. While the field pattern corresponding the $-1$ mode displays a propagating character, the $-1$ mode reveals an exponentially decaying amplitude along the $x_1$ axis in accord with the vanishing diffraction efficiency predicted for the $+1$ mode for large values of the incidence angle.

The dependence of the effective indices associated with the $+1$ and $-1$ SPP modes on the incidence angle $\theta$ provides an additional insight into their nature. In the case of normal incidence the effective indices corresponding to the $+1$ and $-1$ modes belonging to the Au port are nearly identical and have the values $n_{eff}^f(-1)_{\theta = 0^{\circ}} = 0.39$ and $n_{eff}^f(+1)_{\theta= 0^{\circ}} = 0.38$. When the angle of incidence is increased in the range $0 < \theta < 75^{\circ}$ the effective index of the $-1$  mode becomes significantly larger, $n_{eff}^f(-1)_{\theta = 75^{\circ}} = 0.64$, while the effective index of the $+1$  mode becomes smaller $n_{eff}^f(+1)_{\theta = 75^{\circ}} = 0.13$.
%
%

We observe similar, although quantitatively somewhat different, behavior of the transmittance for the SPP eigenmodes belonging to the Al port i.e. propagating in the opposite direction. The transmittance of the  $-1$ mode propagating in the $Al\rightarrow Au$ direction in the case of normal incidence is somewhat larger, $T_b(-1)_{\theta = 0^{\circ}} = 0.16$, than that of the $+1$ mode, $T_b(+1)_{\theta = 0^{\circ}} = 0.09$.
When the angle of the incidence is increased, the difference in the transmittancies of the $+1$ and $-1$ modes increases, and for a certain angle of incidence $\theta \gtrsim 60^{\circ}$ the transmittance of the $+1$ mode vanishes.  Specifically, the transmittance belonging to the $-1$ mode is slightly increased,  $T_f(-1)_{\theta = 75^{\circ}} = 0.17$, when $\theta = 75^{\circ}$, while the transmittance of the $+1$ mode becomes negligible, $T_f(+1)_{\theta = 75^{\circ}} = 0.0006$.

\begin{figure}[h]
\epsfig{file=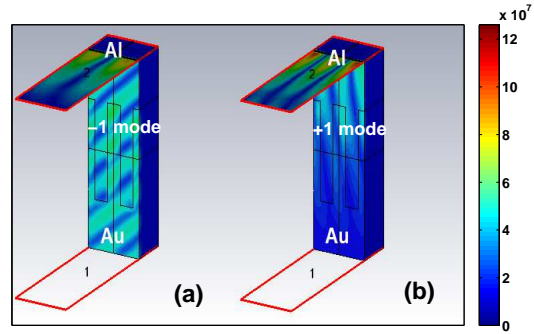,angle=0,width=\columnwidth}
\caption{(Color online) Spatial intensity distributions of the electric fields along the waveguide structure associated with (a) the $-1$ and (b) the $+1$ backward propagating SPP Bragg modes incident at an angle $\theta = 75^{\circ}$. The electric fields are in $Vm^{-1}$ units.}
\label{fig9}
\vspace{0.1cm}
\end{figure}

This behavior is demonstrated in Fig.\ref{fig9} where the spatial distributions of the electric field intensities along the waveguide structure for the $-1$ (a) and $+1$ (b) Bragg beams are shown for the Al waveguide port. Likewise for the forward propagating wave, the field pattern corresponding to the $+1$ mode displays a propagating character, while the $-1$ mode reveals an exponentially decaying amplitude along the $x_1$ axis in accord with the vanishing diffraction efficiency predicted for the $+1$ mode for large values of the incidence angle.
The effective indices associated with the $+1$ and $-1$ SPP modes belonging to the Al port in the case of normal incidence are identical and have values $n_{eff}^b(+1)_{\theta = 0^{\circ}} = n_{eff}^b(-1)_{\theta= 0^{\circ}} = 0.32$. When the angle of incidence is increased in the range $0 < \theta < 75^{\circ}$ the effective index of the $-1$  mode becomes significantly larger, $n_{eff}^b(-1)_{\theta = 75^{\circ}} = 0.58$, while the effective index of the $+1$  mode becomes negligible  $n_{eff}^b(+1)_{\theta = 75^{\circ}} = 0.006$.

The results shown in Figs.\ref{fig7} and \ref{fig9} illustrate the two key effects associated with the behavior of the first-order Bragg modes at oblique incidence. First, one of the first-order beams$(+1)$ is completely suppressed when the SPP impinges the interface at a sufficiently large angle, $\theta \gtrsim 60^{\circ}$. Second, as a result, the structure supports only the $+1$ Bragg beams, with the backward propagating beam having a significantly smaller transmittance than that of the forward propagating one, and yields the contrast transmissivity ratio $R_c = 0.25$.

To compare the results obtained on the basis of the thin phase screen model and those obtained from numerical simulations one has to take into account
the approximation associated with the former approach. Namely, the theoretical model does not take into account radiative modes, and the electric fields associated with the incident and transmitted waves are expressed in terms of a Fourier expansion. On the other hand, the frequency-domain solver deals with a finite set of the eigenmodes supported by the structure,
the majority of which are radiative. The differences in both approaches are reflected in the differences in the quantitative parameters describing an asymmetry in the transmittance. However,one can see that the diffraction efficiencies obtained from the numerical simulations confirm qualitatively the predictions of the thin phase screen model.
 Namely, for a certain magnitude of the modulation depth of the interface between the two metallic regions, which are characterized by a sufficiently large refractive index contrast at a certain wavelength, the transmittance of the zero-order Bragg beam is suppressed while the $+1$ and $-1$ modes become dominant among the surface modes supported by the grating planar structure. We have shown that such a configuration, which supports at normal incidence propagation of both the $+1$ and $-1$ Bragg beams, offers the possibility of  modifying the diffraction efficiency of the $+1$ and $-1$ modes by varying the angle of incidence. Specifically, we found that for a sufficiently large angle of incidence the structure supports only the $-1$ propagating beam, which leads to a substantional asymmetry in the transmittance characterized by the contrast transmissivity ratio $R_c = 0.25$. This value can be increased by carefully choosing both the geometrical and material parameters of the structure.

\bigskip

\section{Conclusion}

We have demonstrated that a system consisting of an air-bimetal interface, whose boundary between the two metallic segments is periodically modulated,  possesses a different transmissivity for a surface plasmon polariton incident on it from one side of it than it has for a surface plasmon polariton incident on it from the opposite side.
This asymmetric transmission of a surface plasmon polariton is based on the suppression of the zero-order Bragg beam, which is not transmitted through the structure for a certain value of the modulation depth of the periodically corrugated boundary between the metals.
Consequently, the mechanism for the asymmetry in the transmittance is related to the higher Bragg modes that are excited on the composite metallic waveguide structure. We have shown that the diffraction efficiencies of the $+1$ and $-1$ Bragg beams can be modified by varying the period and/or the angle of incidence, and for a certain range of the incidence angle one can observe asymmetry in the transmittance of the $-1$ mode while the $+1$ mode is completely suppressed. By varying the material and geometrical parameters of the diffractive structure one can control the contrast transmission that characterizes the degree of the asymmetry.


\section*{Acknowledgements}
The research of V. K. was supported by Grant No. LH 12009 of the Czech Ministry of Education within programme KONTAKT II(LH).


\begin{thebibliography}{99}

\bibitem{Agranovich} {\it Surface Polaritons}, edited by V.M. Agranovich and D. L. Mills (North-Holland, Amsterdam, 1982).


\bibitem{Haldane} F. D. M. Haldane and S. Raghu, Phys. Rev. Lett.  100, 013904 (2008).

\bibitem{Wang} Z. Wang, Y. D. Chong, J. D. Joannopoulos, and M. Soljacic, Phys. Rev. Lett.  100, 013905 (2008).

\bibitem{figotin} A.Figotin and I.Vitebsky, Phys. Rev. E 63, 066609 (2001).

\bibitem{Yu} Z. Yu, G. Veronis, Z. Wang, and S. Fan, Phys. Rev. Lett. {\bf 100} 023902 (2008).

\bibitem{KEV} V. Kuzmiak, S. Eyderman, and M. Vanwolleghem, Phys. Rev. B {\bf 86} 045403 (2012).

\bibitem{KM} V. Kuzmiak and A. A. Maradudin, Phys. Rev. A {\bf 86} 043805(2012).

\bibitem{thin_phase} W. T. Welford, Contemp. Phys. {\bf 21} 401(1980).

\bibitem{Se1} A. E. Serebryannikov, Phys. Rev. B {\bf 80} 155117(2009).

\bibitem{Se2} A. E. Serebryannikov and E. Ozbay, Opt. Express {\bf 17} 13335(2009).

\bibitem{Se3} S. Cakmakyapan, A. E. Serebryannikov, H. Caglayan, and E. Ozbay, Opt. Lett. {\bf 35} 2597(2010).



\bibitem{CST} CST Computer Simulation Technology AG, http://cst.com.

\bibitem{Sheng} P. Sheng, R. S. Stepleman, P. N. Sanda, Phys. Rev. B {\bf 26} 2907(1982).







\end{thebibliography}
\end{document}